# Value–at-Risk and Expected Shortfall for the major digital currencies

Stavros Stavroyiannis[1]

**Abstract**

Digital currencies and cryptocurrencies have hesitantly started to penetrate the investors, and the next step will be the regulatory risk management framework. We examine the Value-at-Risk and Expected Shortfall properties for the major digital currencies, Bitcoin, Ethereum, Litecoin, and Ripple. The methodology used is GARCH modelling followed by Filtered Historical Simulation. We find that digital currencies are subject to a higher risk, therefore, to higher sufficient buffer and risk capital to cover potential losses.

---

[1] Corresponding author, Professor, Department of Accounting & Finance, Technological Educational Institute of Peloponnese, Greece, email: computmath@gmail.com

## 1. Introduction

Digital currencies and cryptocurrencies have started to attract attention from the investors during the last years. Although there are about 800 different implementations that have surfaced, the market capitalization dominance, as of 31-July-2017, is mostly on BitCoin (BTC, or XBT according to the ISO 4217 standard) by 51.14%, followed by Ethereum (ETH) 20.79%, Ripple 7.05% (XRP), and Litecoin 2.47%. Digital currencies, except BTC, have not attracted much attention in the literature. A possible reasoning is that by the end of 2016 BTC was the dominant digital currency with 90% of market capitalization. This year, this dominance has been compromised by the rest of the digital currencies dropping BTC to almost half of the dominance percentage. The price formation of BTC has been studied by a variety of authors (Buchholz et al., 2012; Kristoufek (2013); van Wijk, 2013; Bouoiyour and Selmi 2015; Ciaian et al., 2016; Bouri et al. 2017), as well as, the interconnection between demand and supply (Buchholz et al. 2012; Bouoiyour and Selmi 2015). The main finding of these works is that BTC price is mostly driven by the interaction between supply and demand, and Kristoufek (2013) shows that standard economic theories cannot explain the price formation of BTC. Dyhrberg (2016a; 2016b) examined if BTC possesses some of the hedging abilities of gold, and whether or not it can be included in the variety of tools available to market analysts to hedge market specific risk. The high volatility of BTC price is considered to be the first bubble formation of a virtual currency. This was identified by Phillips et al. (2013), where using the generalized Sup ADF test, detected two bubbles during 2013 and another mini bubble during 2012. It appears that BTC is detached from macro-financial developments, having its own intrinsic risk. The inefficiency of the BTC market has been questioned by Urquhart (2016) and Nadarajah and Chu (2017). Urquhart (2016) working on an earlier sample concludes that BTC in an inefficient market but may be in the process of moving towards an efficient market. On the other hand, Nadarajah and Chu (2017) using an odd power transformation for the BTC returns, they indicate that BTC is actually market efficient.

A critical issue with digital currencies is the fact that there is yet no regulation and no financial instruments. Also, with respect to the initial formal definition of BTC by Satoshi Nakamoto (2008), this peer-to-peer version of electronic cash would allow online payments without going through a financial institution. As a result, digital currencies and cryptocurrencies possess a very shadowy history, including accusations regarding securities theft, fraud, and criminal activity (Frunza, 2016). If an investor has a position in digital currencies then the next logical step is to quantify the risk and the capital requirements needed for future possible losses. Risk management has been the subject of a variety of Accords from the Basel Committee on Banking Supervision (BCBS) (bcbs107b, 2004; bcbs193a, 2011a). Value-at-Risk (VaR) has become a standard tool to quantify market risk, due to its simplicity to capture market risk by a single number. VaR is not subadditive violating the concept of diversification, and presents the problem of aggregation of compartmentalized risk in large financial institutions. Due to the fact that VaR is not a coherent measure of risk, Expected Shortfall (ES) emerged as a natural alternative (Acerbi & Tasche, 2002) fulfilling all four axioms of a coherent risk measure set by Artzner et al. (1997, 1999). This has been taken into account by BCBS (bcbs_wp19, 2011b; bcbs219, 2012) and Basel III Committee agreed to replace the VaR with the Expected Shortfall for the internal model-based approach. In the same vein, the committee had to recalibrate the confidence level for consistency issues. Instead of using the 99% confidence level for the VaR, the Basel III Committee recommends to use the 97.5% confidence level for the ES.

The main contribution of this paper in the existing literature is to quantify for the first time VaR and ES, for BTC, ETH, XRP, and LTC, using Filtered Historical Simulation (FHS), (Barone-Adesi et al., 2002; Giannopoulos and Tunaru, 2005), and to compare the results with the Standard and Poor's S&P500 as a proxy index. The remaining of the paper is organized as follows. Section 2 describes the data and the econometric methodology used, Section 3 presents the results and discusses the findings, and Section 4 concludes.

## 2. Econometric methodology

### 2.1. The data

BTC price data is sourced from Coindesk Price Index (coindesk.com), ETH, XRP, and LTC from coinmarketcap.com, and the Standard and Poor's S&P500 index from finance.yahoo.com, from 8-Aug-2015 to 10-July-2017. All digital currencies trade seven days per week, and the missing values in the rest of the series are filled via sequential linear interpolation for the weekend data, following the procedure in Dyhrberg (2016a; 2016b). The series consist of 704 data entries and the returns are defined via the successive logarithmic differences of the close price.

### 2.2. The model

The dynamics of the AR(1)-GJR(1,1) model (Glosten et al., 1993) is expressed as follows:

$$r_t = \mu + \varphi r_{t-1} + \varepsilon_t = r_t = \mu + \varphi r_{t-1} + \sigma_t z_t \qquad (1)$$

$$\sigma_t^2 = \omega + a\varepsilon_{t-1}^2 + \gamma \varepsilon_{t-1}^2 I(\varepsilon_{t-1} < 0) + \beta \sigma_{t-1}^2 \qquad (2)$$

where $a$ and $\beta$ are the ARCH and GARCH coefficients, $\gamma$ is the leverage effect capturing the asymmetry effect in return volatility, and $I$ is an indicator function taking the value 1 when $\varepsilon_{t-1} < 0$ and zero otherwise, and the residuals follow the standardized Pearson type-IV (PIV) distribution (Stavroyiannis et al., 2012),

$$f(z) = \frac{\hat{\sigma}\Gamma\left(\frac{m+1}{2}\right)}{\sqrt{\pi}\Gamma\left(\frac{m}{2}\right)} \left|\frac{\Gamma\left(\frac{m+1}{2} + i\frac{v}{2}\right)}{\Gamma\left(\frac{m+1}{2}\right)}\right|^2 \frac{\exp(-v\tan^{-1}(\hat{\sigma}z + \hat{\mu}))}{(1 + (\hat{\sigma}z + \hat{\mu})^2)^{\frac{m+1}{2}}} \qquad (3)$$

$$\hat{\mu} = -\frac{v}{m-1} \qquad (4)$$

$$\hat{\sigma} = \sqrt{\frac{1}{m-2}\left(1 + \frac{v^2}{(m-1)^2}\right)} \qquad (5)$$

where $\Gamma(\cdot)$ is the Gamma function and $i$ the imaginary unit. All programming has been performed with the Matlab, MathWorks® computing language.

## 3. Results of the econometric methodology

### 3.1. Descriptive statistics and stylized facts

The descriptive statistics of the returns are shown in Table 1. All series exhibit statistically significant skeweness and kurtosis, and the Jarque-Bera test shows deviation from normality. The ARCH test shows that all series exhibit heteroskedastivity. The Lung-Box test on the squared residuals is statistically significant for all series and shows the presence of autocorrelation. The returns show autocorrelation for the XRP and LTC. A striking difference of the virtual currencies with respect to S&P500 index is the one order of magnitude higher return and standard deviation.

**Table 1 Descriptive statistics of the return series.**

|        | mean   | std.  | skew.   | kurt.  | J.B.   | ARCH(12) | LB(12) | LB-2(12) |
|--------|--------|-------|---------|--------|--------|----------|--------|----------|
| S&P500 | 0.0002 | 0.007 | -0.253* | 9.245* | 1150.* | 12.53*   | 14.39  | 214.3*   |
| BTC    | 0.0031 | 0.031 | -0.979* | 8.445* | 980.7* | 7.549*   | 8.090  | 109.06*  |
| ETH    | 0.0081 | 0.075 | 0.685*  | 7.780* | 723.2* | 7.904*   | 20.39  | 146.61*  |
| XRP    | 0.0045 | 0.074 | 4.183*  | 67.80* | 1.e+5* | 8.006*   | 48.52* | 114.34*  |
| LTC    | 0.0036 | 0.050 | 2.378*  | 24.41* | 14088* | 4.036*   | 28.94* | 51.818*  |

*Notes: J.B. is the statistic for the null of normality; ARCH(12) denotes the test for heteroskedasticity, LB(12) denotes the Ljung–Box test statistic for serial correlation, LB-2(12) denotes the Ljung–Box test statistic for serial correlation on the squared residuals with 12 lags respectively. (\*) denotes statistical significance at the 5% critical level.*

The correlations of the time series returns are shown in Table 2.

**Table 2 Correlations of the time series returns**

|        | S&P500  | BTC     | ETH     | XRP     | LTC    |
|--------|---------|---------|---------|---------|--------|
| S&P500 | 1       | -0.0292 | -0.0080 | -0.0130 | 0.0045 |
| BTC    | -0.0292 | 1       | 0.14801 | 0.5298  | 0.1054 |
| ETH    | -0.0080 | 0.14801 | 1       | 0.0949  | 0.0158 |
| XRP    | -0.0130 | 0.52982 | 0.0949  | 1       | 0.1745 |
| LTC    | 0.0045  | 0.10539 | 0.0158  | 0.1745  | 1      |

The correlations of the digital currencies with the S&P500 proxy market is practically zero. The highest correlation is observed for the BTC-XRP pair (0.53), and the lowest correlation between the ETH-LTC pair (0.016). Ranking the rest of the correlations in a descending order we get BTC-ETH (0.15), XRP-LTC (0.17), BTC-LTC (0.10), and ETH-XRP (0.095).

### 3.2. Univariate GARCH results

The results of the univariate GARCH methodology are shown in Table 3.

**Table 3** Results of the univariate GJR(1,1) model.

|   | S&P500 | BTC | ETH | XRP | LTC |
|---|---|---|---|---|---|
| $\mu$ | 2.2e-04 | 0.0024* | 0.0031 | -0.0016 | 0.0011 |
| $\varphi$ | 0.0845* | -0.0911* | 0.0279 | -0.0035 | -0.0936* |
| $\omega$ | 2.3e-05* | 1.3e-05 | 2.6e-04 | 2.3e-04 | 2.1e-05 |
| $a$ | 0.0000 | 0.2665* | 0.2820* | 0.5051 | 0.1855* |
| $\beta$ | 0.9065* | 0.8488* | 0.6983* | 0.5562* | 0.8937* |
| $\gamma$ | 0.1903* | -0.2232* | 0.0429 | -0.1296 | -0.1679* |
| $\nu$ | -0.0943 | 0.2439 | -0.7575* | -0.3397* | -0.3125* |
| $m$ | 2.5570* | 3.2421* | 3.7934* | 2.7649* | 2.5237* |

*(\*) denotes statistical significance at the 5% critical level.*

The parameters $a$ and $\beta$ reflect the short run dynamics of the volatility. The ARCH coefficient $a$ is statistically significant for BTC, ETH, and LTC meaning that the volatility reacts to quite intensively to market movements. The GARCH coefficient $\beta$ is statistically significant for all series, an indication that a shock to the conditional variance takes time to die-out. The leverage coefficient is statistically significant for S&P500, BTC, and LTC. The standardized residuals and the squared standardized residuals of the series do not possess any remaining autocorrelation, which is an important issue in order to apply FHS.

### 3.3. VaR and Expected Shortfall

A general approach to VaR and ES is either a calculation based on the historical simulation approach, ignoring the structure and distribution of the returns, or using Monte Carlo simulation on a parametric model for variance, incorporating a sufficiently large sample of random numbers from a specific distribution. The FHS methodology is a combination of both approaches via bootstrapping on the existing standardized residuals. After filtering out most of stylized facts via the AR(1)-GJR(1,1) model, the VaR for a specific confidence level $\theta$ is the quantile that solves the equation

$$VaR(1-\theta) = -inf\{q \in R | F_{PIV}(x) \geq q\} \qquad (6)$$

where $F_{PIV}$ is the cumulative distribution of Eq.(3). The VaR levels for the long position for the $\theta$ confidence level are identified as

$$VaR(long) = \mu_t + F_{PIV}^{-1}(1-\theta)\sigma_t \qquad (7)$$

where $F_{PIV}^{-1}$ is the inverse of the cumulative distribution function of Eq. (3) at the specific confidence level. The FHS is implemented as follows; firstly, the volatility of the asset is modeled via GARCH methodology as in Eq. (2), and the standardized residuals $z_t = \varepsilon_t/\sigma_t$ are computed. At this point an assumption about the distribution of the returns has to be made, and the PIV distribution is chosen. Secondly, instead of drawing random numbers from a specific distribution, the samples are drawn with replacement from the computed standardized residuals. Thirdly, using Eqs(1-2) the hypothesized returns are constructed, and VaR and ES are calculated. For the FHS we consider 100000 trials, for a horizon of 10 trading days. ES is calculated via the Arcebi and Tasche (2002) approach,

$$ES(1-\theta) = -\frac{1}{[(1-\theta)N]} \sum_{i=1}^{[1-\theta)N]} X_i^\dagger \qquad (8)$$

where the dagger symbol (†) denotes the ascending order sorting of the FHS, and [·] is the integer part of the VaR level times the number of simulations. In agreement with the Basel III committee, since we report on both the VaR and ES, the results for VaR and ES for the long position are shown in Table 4, for $VaR = 1 - \theta \in$ [0.10, 0.05, 0.025, 0.01] in percentages.

**Table 4 VaR and Expected Shortfall**

|            | S&P500(%) | BTC(%) | ETH(%) | XRP(%) | LTC(%) |
|---|---|---|---|---|---|
| VaR(0.10)  | 2.2107    | 5.5529 | 21.609 | 19.632 | 23.136 |
| ES(0.10)   | 4.3368    | 11.469 | 37.636 | 34.535 | 40.136 |
| VaR(0.05)  | 3.3170    | 9.3023 | 31.296 | 27.712 | 33.713 |
| ES(0.05)   | 5.9861    | 15.774 | 49.479 | 45.972 | 52.530 |
| VaR(0.025) | 4.6509    | 13.445 | 42.132 | 37.004 | 45.370 |
| ES(0.025)  | 8.0838    | 20.460 | 62.936 | 60.227 | 66.209 |
| VaR(0.01)  | 7.1706    | 19.239 | 59.152 | 52.057 | 61.299 |
| ES(0.01)   | 11.790    | 27.247 | 83.592 | 86.036 | 87.232 |

It is clear that the capital requirements of the digital currencies are higher that the S&P500 proxy market; however, BTC, due to the former market domination, appears to behave more stable than the other three digital currencies. As a next step of our analysis we test whether the aforementioned hypothesis of the Basel III committee holds for digital currencies. The hypothesis is that the quantity of risk measured by a 99% VaR is approximately the same as a 97.5% ES. Looking at Table 4, we can confirm that the 99% VaR (the penultimate line) is almost equal to the 97.5% ES (third line from the end) for BTC and ETH, while XRP and LTC hold a larger gap.

## 4. Conclusion

This paper estimates the VaR and ES risk measures for a horizon of 10 days for the major digital currencies. An AR(1)-GJR(1,1) framework is specified to model the stylized facts of the autocorrelation of the returns, volatility clustering, and the asymmetry effect, and the risk measures are calculated using FHS. BCBS strives to strengthen the financial system against systemic risks and this will have implications on risk management and the cost of capital and liquidity. The capital adequacy rules set a minimum requirement on the size of the financial buffer based on the assumed risk. The financial system after the last crisis is susceptible to changes, and a controversial issue is whether the new digital currencies and cryptocurrencies will be regulated as new payments systems. Until these issues accomplish, an investor should be aware that a long position in digital currencies is subject to a much higher risk than proxy markets; therefore, to higher sufficient buffer and risk capital to cover potential losses.